\documentclass[fleqn,twoside]{article}
\usepackage{espcrc2}

\usepackage{graphicx}
\usepackage[figuresright]{rotating}

\newcommand{\AmS}{{\protect\the\textfont2
  A\kern-.1667em\lower.5ex\hbox{M}\kern-.125emS}}

\title{Gribov Copies and Gauge Fixing in Lattice Gauge Theories}

\author{O. Oliveira\address[MCSD]{Departamento F{\'\i}sica, 
                                  Universidade de Coimbra \\
                                  3004-516 Coimbra, Portugal}
        \thanks{orlando@teor.fis.uc.pt},
        P. J. Silva\addressmark\thanks{paulojesussilva@mail.telepac.pt}
        }
       
\begin{document}

\begin{abstract}
We address the problem of the gauge fixing versus Gribov copies
in lattice gauge theories. For the Landau gauge, results show that a suitable
combination of evolutionary algorithms with traditional steepest descent
methods identifies the global maximum of the optimisation function. We discuss
the performance of the combined algorithm on small cubic lattices for SU(2)
and SU(3).
\vspace{1pc}
\end{abstract}

\maketitle

\section{Introduction and Motivation}

The formulation of gauge theories on the lattice does not require gauge fixing.
However, to study Green's functions of the fundamental fields, it is
unavoidable to pick a gauge. Moreover, the correlation functions of the
fundamental fields can be used to compute renormalization constants or 
non-perturbative coefficients used in numerical simulations. For a general
overview on gauge fixing in lattice gauge theories and related issues see
\cite{Giu01}.

On the lattice, gauge fixing can be viewed as an optimisation problem.
Typically, we have a maximising function with many local maxima, the Gribov
copies, and we want to identify its global maximum. The global maximum is not
always unique defined. From the numerical point of view, global optimisation
is not a trivial task, and among the various techniques Genetic Algorithms
(GA) \cite{Mich96} seems to be a good global searching method. 

Genetic Algorithms require an evolutive population and assume a number of
rules for reproduction. The tuning of basic parameters, the probabilities
of the genetic operators and the size of the evolutive population, is crucial
to have good performance. The size of the
population is an important parameter and can easily drive the method to use
as much memory as available.

Our investigation started using pure GA to SU(3) and SU(2) Landau gauge
fixing on $4^4$ and $8^4$ lattices. We conclude that the method
was unable to identify conveniently the global maximum. This negative result is
due to the large number of variables involved and to the structure of the
optimisation function. Fortunately, a combination of GA with steepest descent
method (SD) \cite{Dav88} seems to be able to converge to the global maximum.

\begin{figure}[h]
%\vspace{9pt}
%\framebox[55mm]{\rule[-21mm]{0mm}{43mm}}
\scalebox{0.5}[0.3]{\includegraphics{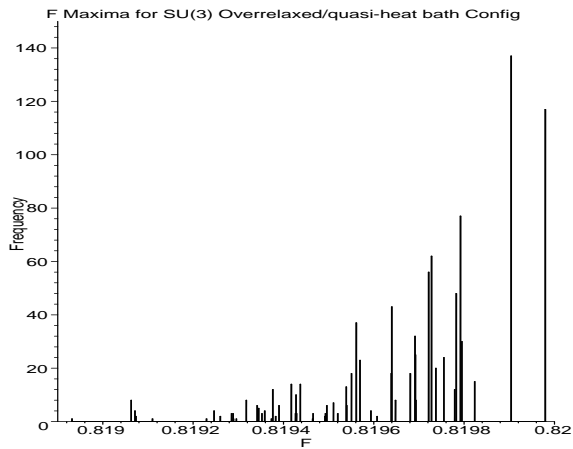}}
\caption{Local maxima from an $8^4$ SU(3) configuration obtained after 1000 SD 
starting at randomly chosen different points.}
\label{maxima}
\end{figure}

\section{Evolution Program and Landau Gauge Fixing}

Landau gauge fixing is implemented by global maximising the function of
the links
\begin{equation}
   F[U] \, = \, \sum\limits_{x,\mu} \, \mbox{Re} \left\{ \mbox{Tr} \left[
                g(x) \, U_\mu (x) \, g^\dagger (x + \mu)
                \right) \right\}
\label{fit0}
\end{equation}
over the gauge orbits.

\begin{figure}[t]
%\framebox[55mm]{\rule[-21mm]{0mm}{43mm}}
\scalebox{0.31}[0.3]{\includegraphics{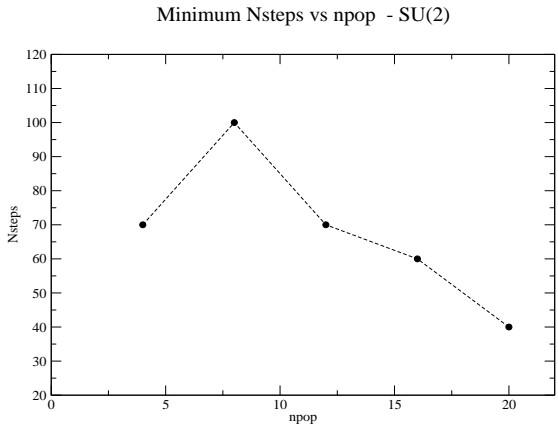}}
\caption{$N_{\mbox{steps}}$ as function of $npop$ for a random SU(2) 
configuration.}
\label{npopnstepssu2}
\end{figure}

Our implementation of the combined GA and SD algorithm uses the genetic
operators (after reunitarisation): 
$i)$ random crossover, $ii)$ random blending, 
and the mutation operators
\begin{eqnarray}
 & & g(x)  \longrightarrow  g(x) \, + \epsilon A , \\
 & & g(x)  \longrightarrow  A ,  \\
 & & g(x)  \longrightarrow    g(x) \left( 1 \, + \epsilon A \right),
\end{eqnarray}
where $\epsilon$ ($\left| \epsilon \right| \le 0.025$) is a random number and
$A$ is a random SU(N) matrix.

As fitness function we take
\begin{equation}
 \widetilde{F} [g] \, = \, N_{\mbox{steps}} 
 \,\, \mbox{ SD iterations on }  \,\,
  F[U^g] \, ,
  \label{fitness}
\end{equation}
i.e. the fitness function is obtained from (\ref{fit0}) 
after $N_{\mbox{steps}}$ steepest descent iterations.

The code starts by generating an uniformly distributed initial population 2.5
times larger than the evolutive population. 
The first generation is chosen from this initial population. The mating 
selection for reproduction, and the choice of the first evolutive generation, 
uses roulette wheel sampling favouring the best members of the population. 

On the final generation we apply SD requiring 
$\left| \partial A \right| < 10^{-10}$ to all members of the population.

\begin{figure}[t]
%\framebox[55mm]{\rule[-21mm]{0mm}{43mm}}
\scalebox{0.31}[0.3]{\includegraphics{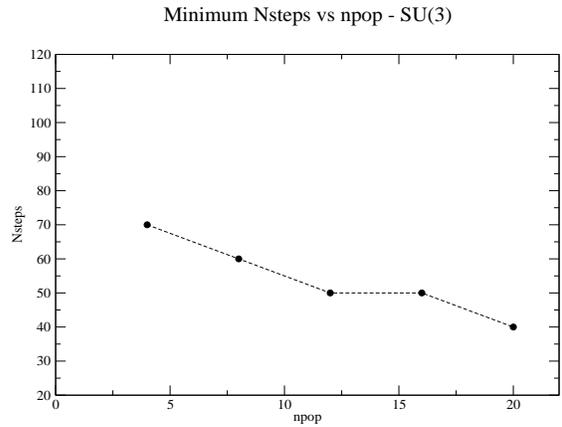}}
\caption{$N_{\mbox{steps}}$ as function of $npop$ for $\beta = 5.7$ SU(3) 
configurations.}
\label{npopnstepssu3}
\end{figure}

\begin{figure}[h]
%%\framebox[55mm]{\rule[-21mm]{0mm}{43mm}}
\scalebox{0.31}[0.3]{\includegraphics{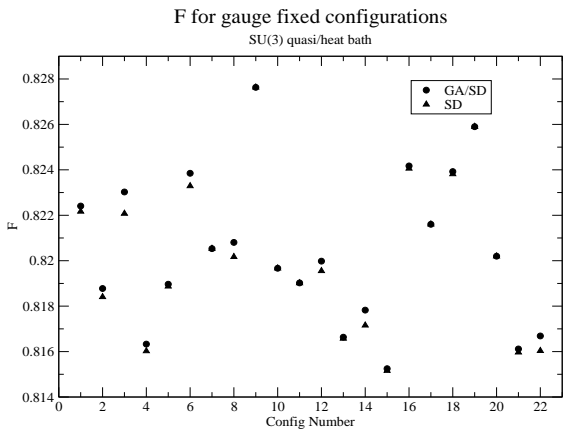}}
\caption{$F$ for gauge fixed SU(3) configurations.}
\end{figure}
\begin{figure}[h]
%\framebox[55mm]{\rule[-21mm]{0mm}{43mm}}
\scalebox{0.31}[0.3]{\includegraphics{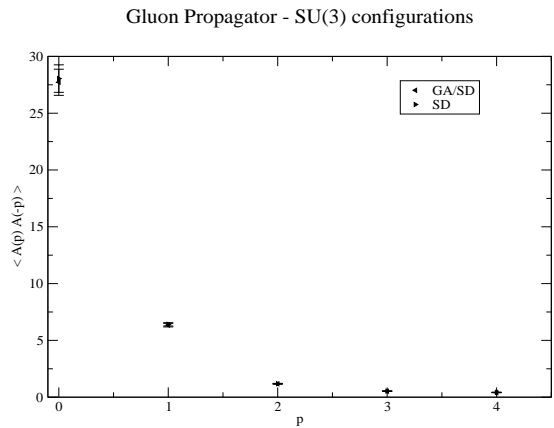}}
\caption{SU(3) gluon propagator.}
\label{propagador}
\end{figure}

\section{Results and Discussion}

The combined algorithm was tested with SU(2) $4^4$ random configurations, and
$8^4$ ($\beta \, = \, 5.7$) SU(3) configurations generated with version 6 of 
the MILC code \cite{MILC}.

All runs report results after 400 GA generations. Concerning CPU time we
are not yet in the position on giving numbers. This is because we performed a
large number of convergence tests at intermediate generations. Final results,
including more details about the method and CPU time information will be
reported elsewhere soon. 

For 3 of the SU(3) configurations and for the SU(2) configurations,
the global maximum was computed by performing a large number (1000) of SD
starting from randomly chosen different points. In our case, this should not
be a problem since we are working with relatively small lattices. 
If we compare the results of the combined algorithm with random search,
we have never observed maxima larger than that obtained by random search. 

Configurations show a large number of local maxima - see figure \ref{maxima}.
For the combined algorithm, the convergence of the best member of the
population to the global maxima of $F$ depends on the size of the evolutive
population ($npop$) and on $N_{\mbox{steps}}$. 
Although one can arrive at the proper answer with quite small
evolutive populations, the number of single steepest descent iterations 
defining the fitness function (CPU time) increases as the size of the
population decreases. In figures \ref{npopnstepssu2} and \ref{npopnstepssu3}
one can see the relation between $npop$ and $N_{\mbox{steps}}$ for SU(2) and
SU(3) configurations.

Results are encouraging but further studies of how the algorithm scales
with volume and $\beta$ should be done. The tests were performed using a 
serial version of the code. A parallel implementation of the algorithm is
about to start testing.

To address the problem of Gribov copies we look at the simplest correlation
function which can be computed, namely the gluon propagator. In order to
compute the propagator, 22 over-relaxed/quasi heat bath SU(3) configurations 
where generated and
the correlation function from SD compared to the correlation function
from the combined GA/SD algorithm. Results are given in figure
\ref{propagador}. 
The two methods show essentially the same
behaviour. However, due to the large number of maxima in some configurations,
a careful analysis should be done before drawing conclusions. Furthermore,
if an analysis of the gluon propagator certainly must be done, investigations
of the quark and ghost propagators should not be forgotten.

\vspace{0.5cm}
Work supported by PRAXIS/C/FIS/14195/98.

\end{document}